\newif\ifPDFLaTeX
\expandafter\ifx\csname pdfoutput\endcsname\relax\else\PDFLaTeXtrue\fi
 \documentclass[a4paper,12pt]{article}

\ifPDFLaTeX
  \usepackage{type1cm}
  \usepackage[pdftex,colorlinks]{hyperref}
  \usepackage[pdftex]{graphicx,feynmp}
\else
 \usepackage{graphicx,feynmp}
\fi

\usepackage{amsmath,amssymb}
\usepackage{url}
\usepackage{cite}
\providecommand{\preprintno}[1]{\relax}
\usepackage{fancyhdr}
\def\preprint#1{\gdef\thepreprint{#1}}
\def\thepreprint{}
\fancypagestyle{fancyplain}{
  \fancyhf{}
  \fancyhead[R]{\thepreprint\\ \hfill\\ January 7, 2005}
   
  \addtolength{\headsep}{3.5cm}
  \addtolength{\headheight}{1cm}
}
 \makeatletter
   \def\fps@figure{t}
   \def\fps@table{b}
   \long\def\@makecaption#1#2{%
    \vskip\abovecaptionskip
     \sbox\@tempboxa{#1: \textit{#2}}%
     \ifdim\wd\@tempboxa>\hsize
       #1: \textit{#2}\par
     \else
       \global\@minipagefalse
       \hb@xt@\hsize{\hfil\box\@tempboxa\hfil}%
     \fi
     \vskip\belowcaptionskip}
 \makeatother
\makeatletter
\renewcommand\section{\@startsection {section}{1}{\z@}%
                                   {-3.5ex \@plus -1ex \@minus -.2ex}%
                                   {2.3ex \@plus.2ex}%
                                   {\large\bf}}   
\renewcommand\subsection{\@startsection {subsection}{1}{\z@}%
                                   {-3.5ex \@plus -1ex \@minus -.2ex}%
                                   {2.3ex \@plus.2ex}%
                                   {\normalfont\bf}}         
\makeatother
\allowdisplaybreaks
\newcommand{\ii}{\mathrm{i}}

\makeatletter
\newcommand{\fmslash}[2][0mu]{%
  \mathchoice
    {\fmsl@sh\displaystyle{#1}{#2}}%
    {\fmsl@sh\textstyle{#1}{#2}}%
    {\fmsl@sh\scriptstyle{#1}{#2}}%
    {\fmsl@sh\scriptscriptstyle{#1}{#2}}}      
\newcommand{\fmsl@sh}[3]{%
  \m@th\ooalign{$\hfil#1\mkern#2/\hfil$\crcr$#1#3$}}
\makeatother
\begin{document}
%%%%%%%%%%%%%%%%%%%%%%%%%%%%%%%%%%%%%%%%%%%%%%%%%%%%%%%%%%%%%%%%%%%%%%%%
\begin{fmffile}{top_ecfapics}
\fmfset{arrow_len}{3mm}
%%%%%%%%%%%%%%%%%%%%%%%%%%%%%%%%%%%%%%%%%%%%%%%%%%%%%%%%%%%%%%%%%%%%%%%%

\preprint{MZ-TH/04-21\\ LC-TH-2004-030\\ hep-ph/0412028}
\title{ Top quark couplings at ILC:\\
 six and eight fermion final states\thanks{Extended notes for a talk given at the 2nd Workshop of the ECFA Study `Physics and Detectors for a Linear Collider' , 1.-4.9 2004, Durham}}
\author{ Christian Schwinn\thanks{schwinn@thep.physik.uni-mainz.de}\\
{\normalsize\it Institut f\"ur Physik, Johannes-Gutenberg-Universit\"at}\\
{\normalsize\it Staudingerweg 7,  D-55099 Mainz, Germany}}
\setcounter{tocdepth}{2}
\setcounter{secnumdepth}{3}
 \date{}

\maketitle
\thispagestyle{fancyplain}
\setcounter{page}{0}
\begin{abstract}
  I discuss the calculation of cross sections for processes with six
  and eight fermions in the final state, contributing to single-top
  production and associated top-Higgs production at a linear collider.
  I describe the schemes for the treatment of finite decay widths
  implemented in the matrix element generator \texttt{O'Mega} and give a
  numerical comparison for single-top production.  In the case of
  single top production, after reducing vector boson fusion
  backgrounds by appropriate cuts, the effect of including the full
  six fermion final state amounts to $2-5\%$.  In associated top-Higgs
  production, non-resonant electroweak backgrounds are of a similar
  magnitude while QCD backgrounds are much larger.
  \end{abstract}
\newpage
\section{Introduction}
Because of its large mass close to the electroweak scale,
the top quark plays a special role in many new physics
models.
Therefore,  determining the
couplings of the top quark  is  an
important goal of future collider experiments in order to distinguish
the minimal standard model from one of its
extensions~\cite{Glover:2004cy}.

The CKM matrix element $V_{tb}$ is presently constrained indirectly
assuming unitarity of the CKM matrix and the expected precision of a
direct measurement at the LHC is $\sim 7\%$. At an international
linear
collider~(ILC)~\cite{Aguilar-Saavedra:2001}, a
comparable precision can be achieved in single top production $e^+ e^-
\to e^-\bar\nu_\mu t\bar b$~\cite{Boos:1996vc,Boos:2001sj} while a
significant improvement requires the $\gamma e^-$ option of a linear
collider~\cite{Boos:2001sj}. An indirect determination is  possible
from the top width measurement at the top-pair production threshold at
ILC~\cite{Martinez:2002st}.

Assuming a Higgs boson is found at LHC, measuring its Yukawa coupling
to the top quark---that is predicted to be equal to $g_{Ht\bar
  t}=m_t/v$ in the standard model on tree level---will be important to
establish its properties and identify the underlying mechanism of
electroweak symmetry breaking.  At LHC, the top quark Yukawa coupling
can be determined to a precision of $\sim 15\%$ if the branching
ratios of a standard model Higgs boson are assumed or if the branching
ratios are measured at an ILC operating at $\sqrt s=500$
GeV~\cite{Desch:2004kf}.  At an ILC with $\sqrt s =800-1000$ GeV, in
associated top Higgs production~\cite{Djouadi:1992gp} $e^+ e^-\to
t\bar t H$ a precision of $\sim 5-10\%$ can be reached in the
determination of the top quark Yukawa
coupling~\cite{Baer:1999,Juste:1999}, depending on the value of the
Higgs mass. Again,
an indirect measurement of $g_{Ht\bar t}$ is possible
at the top-pair production threshold~\cite{Martinez:2002st}.

In this note, I describe the calculation of tree level electroweak
cross sections for six and eight fermion final states, contributing to
single top production and associated top-Higgs production,
respectively.  The calculations have been performed using the matrix
element generator \texttt{O'Mega}~\cite{OMega} and the adaptive
Monte Carlo phase space and event generator
\texttt{WHI\-ZARD}~\cite{Kilian:WHIZARD}.  Results for QCD backgrounds to
associated top-Higgs production obtained with
\texttt{MadGraph}~\cite{Stelzer:1994ta} are also presented.  It is pointed out
that a consistent treatment of finite width effects is essential in
obtaining numerically reliable predictions.  Similarly, violating
gauge invariance by including only subsets of Feynman diagrams can
give wrong results in important regions of phase space.

In section~\ref{sec:omega}, I give a brief update on \texttt{O'Mega} and
\texttt{WHI\-ZARD}, including a description of schemes implemented for the
treatment of unstable particles.  Single top production and the
relevant backgrounds from top-pair production and vector boson fusion
are discussed in section~\ref{sec:single-top}.  Associated top Higgs
production, backgrounds from associated top-$Z$ production and gluon splitting
and the  numerical effects of including the full 
eight fermion final state are discussed in section~\ref{sec:yukawa}.

\section{O'Mega \& WHIZARD}
\label{sec:omega}
Theoretical predictions for the
 physical six particle final states for single
top production require the calculation of cross sections with hundreds
to thousands contributing diagrams.
For this purpose various  computer programs are available and
good agreement among the different codes has been
found~\cite{Moretti:2004je,Dittmaier:2002,Gleisberg:2003bi}.
In associated top Higgs
production, eight fermion final states
with over twenty thousand diagrams appear for the decay mode $H\to b\bar b$. 
First results for such processes 
obtained with \texttt{HELAC/PHEGAS}~\cite{Kanaki:2000ey}
and \texttt{O'Mega/WHI\-ZARD} have been presented~\cite{Avh,Kilian:2004} but a
comprehensive study remains to be performed.  Here, I briefly review
the current status of the programs \texttt{O'Mega/WHI\-ZARD} used in the
calculations described in this note.

In the \texttt{O'Mega} algorithm~\cite{OMega}, the amplitude is
expressed in terms of sub-amplitudes with one external off-shell
particle that can be constructed recursively.  In this construction,
Feynman diagrams are not generated separately and sub-amplitudes
appearing more than once in the amplitude are factorized by
construction, avoiding redundant code.  The one-particle off shell
sub-amplitudes satisfy simple Ward Identities, allowing for
comprehensive gauge checks~\cite{Schwinn:2003} that have be used to
verify the implementation of the Feynman rules of the electroweak
standard model.  \texttt{O'Mega} allows to generate \texttt{fortran}
code for helicity amplitudes with arbitrary external particles, where
masses are treated exactly. The complete electroweak standard model in
unitarity and $R_\xi$ gauge is available, including CKM mixing.  In
addition, Majorana fermions are supported~\cite{Reuter:2002} and the
complete minimal supersymmetric standard model has been implemented.
While QCD interaction vertices are included, the implementation of
interfering color amplitudes is not yet completed.  Thus QCD effects
can presently only be included in simple situations where an overall
color factor is sufficient.  Recent versions of \texttt{O'Mega} allow
to treat cascade decays of unstable particles, either by using the
narrow width approximation or by selecting the diagrams with decay
topology but keeping the decaying particle off-shell.

For the phase space integration and event generation the adaptive
multi-channel Monte Carlo package \texttt{WHI\-ZARD}~\cite{Kilian:WHIZARD} has
been used. In the current version the treatment of multi-particle
final states with identical particles has been improved and good
numerical agreement has been found~\cite{Kilian:2004} between the
multi purpose programs \texttt{O'Mega/WHI\-ZARD} and the dedicated six fermion 
production program
\texttt{LUSIFER}~\cite{Dittmaier:2002}.
For processes with few identical particles in the final state, 
\texttt{O'Mega/WHI\-ZARD} has been found to be more efficient~\cite{Kilian:2004},
for processes with many identical particles the dedicated
program.  
For up to six particles in the
final state also matrix elements generated with
\texttt{MadGraph}  can be used in \texttt{WHI\-ZARD} where the
version currently implemented allows to include QCD effects at a fixed
order of the coupling constant.
 
To obtain correct and numerically stable predictions for cross
sections, it is important to use a consistent scheme for the decay
widths of unstable particles.  
Several schemes have been implemented in  \texttt{O'Mega}
 and compared in
single $W$-production~\cite{Schwinn:2003}.
Here I briefly discuss 
schemes for the top quark propagator~\cite{Kauer:2001}, similar
remarks apply to the treatment of the gauge boson widths.  A numerical
comparison  is given
in~section~\ref{sec:single-top}.
While it is physically sensible to include a finite width only in 
resonant propagators, the resulting \emph{step width} scheme
is in general inconsistent with gauge invariance.
In the \emph{fixed width scheme}, in \texttt{O'Mega}  the prescription
\begin{equation}\label{eq:top-cm}
  S_t=\frac{\ii}{\fmslash p-\sqrt {m_t^2-\ii m_t\Gamma_t}}
 =\frac{\ii (\fmslash p+\sqrt{m_t^2-\ii m_t\Gamma_t})}{p^2-m_t^2+\ii m_t\Gamma_t}
\end{equation}
is used that is consistent with QED gauge invariance, in contrast to a
naive Breit Wigner propagator, i.e.  $S_t=\ii(\fmslash
p-m_t)/(p^2-m_t^2+\ii m_t\Gamma_t)$.  While the fixed width scheme
does not respect $SU(2)$ invariance, in the examples considered
previously (e.g. in~\cite{Dittmaier:2002}) no numerical
inconsistencies have been found.  However, for forward scattering of the electron in the process $e^+e^-\to
b\bar b\mu^-\bar\nu_\mu e^+\nu_e $ considered in
section~\ref{sec:single-top} the fixed width scheme becomes
numerically unstable in Feynman gauge.  In the \emph{complex mass}
scheme~\cite{Denner:1999}, $SU(2)$ gauge invariance is restored by
replacing $ m_t\to \sqrt{m_t^2-\ii m_t\Gamma_t}$ not only in the
propagator but everywhere in the lagrangian, i.e.  also in the
top-Yukawa couplings to the Higgs and Goldstone bosons.  Similar
replacements are performed for the gauge boson masses, leading e.g. to
a complex Weinberg angle.  The complex mass scheme is consistent for
scattering amplitudes where only stable particles appear as external
states, so the application in the single top production process
$e^+e^-\to t\bar b e^-\bar\nu_e$ requires to consider complete six
fermion final states.  Another fully gauge invariant scheme is the
\emph{fudge factor} or overall scheme, where the amplitude is
calculated with vanishing widths and multiplied with an overall factor
$(p^2-m^2)/(p^2-m^2+\ii m\Gamma)$ for each resonant propagator.  The
triple gauge boson vertices in the nonlocal effective Lagrangian
scheme of~\cite{Beenakker:1999} are also available in \texttt{O'Mega}.  A
problematic high energy behavior has been found in simple versions of
this scheme both in the second reference of~\cite{Beenakker:1999} and
in~\cite{Schwinn:2003}. While the implementation of the $Wtb$ vertex
in this scheme into \texttt{O'Mega} is planned for the future, it has not been
used in the calculations described in this note.

\section{Single Top Production}
\label{sec:single-top}
In this section, I discuss cross sections for six fermion final states
contributing to single top production at a linear collider.  An
analysis of the measurement of the CKM matrix element using the
process $e^+ e^-\to e^-\bar\nu_e t\bar b$ has been given
in~\cite{Boos:1996vc} and extended to anomalous couplings and the
$\gamma\gamma$, $e^-\gamma$ and $e^-e^-$ options of a linear collider
in~\cite{Boos:2001sj}.  No six fermion final states have been
considered, however.  Results for the reactions $e^+ e^-\to t \bar b
f\bar f'$ have appeared also in~\cite{Biernacik:2001mp}.  Existing
studies of six fermion final states for top
production~\cite{Accomando:1998,Kolodziej:2001xe,Dittmaier:2002,Gleisberg:2003bi}
have focused on top pair production and found background contributions
of $\sim 5\%$ for $\sqrt s =500$ GeV, becoming more important with
growing center of mass energy. Also distortions in angular and
invariant mass distributions by irreducible backgrounds have been observed.
The impact of an anomalous $W tb$ coupling on 
the angular distributions of the leptonic
decay products of the top quarks in
six fermion final states
for top pair production has been found to be small~\cite{Kolodziej:2003gp}.

The results obtained with \texttt{O'Mega/WHI\-ZARD} have been compared
with existing results given in the literature.  For the top production
process $e^+e^-\to t\bar b e^-\bar\nu_e$ agreement has been found with
the results of~\cite{Biernacik:2001mp} within the errors of the Monte
Carlo integration.  In table~\ref{tab:6f-amegic}, for some six fermion
final states $e^+e^-\to b\bar b e^-\bar \nu_e f_i\bar f_j$ we compare
with the results of~\cite{Gleisberg:2003bi} obtained with 
\texttt{AMEGIC}~\cite{Krauss:2001iv}, adopting the same set of input
parameters and cuts and treating all fermions as massive.  For
comparison, the results of~\cite{Dittmaier:2002} for massless fermions
are also shown.
\begin{table}[htbp]
  \begin{center}
    \begin{tabular}{|c|c|c|c|}
\hline
& \multicolumn{3}{c|}{ $ \sigma_{\text{EW}}(e^+e^-\to b\bar b e^-\bar \nu_e f_i\bar
  f_j)$(fb)}\\
\hline
 $f_i\bar f_j$ & \texttt{O'Mega/WHIZARD} &\texttt{AMEGIC}~\cite{Gleisberg:2003bi}  &\texttt{LUSIFER}~\cite{Dittmaier:2002}\\\hline
 $\mu^+\nu_\mu$ & 5.831 (10)    &5.865 (24) &5.819 (5)\\\hline
$e^+\nu_e$  & 5.871 (12) &5.954 (55)&  5.853 (7)\\\hline
$ u\bar d$ &17.251 (30) & 17.366 (68)& 17.187 (21) \\\hline
\end{tabular}
\caption{Comparison of cross sections for six fermion final states contributing to single top production for $\sqrt s= 500$ GeV }
  \end{center}
\label{tab:6f-amegic}
\end{table}
Only the electroweak contributions are included.  QCD
corrections have been found to increase the cross sections for
semileptonic final states by about
$1\%$ for $\sqrt s =500$ GeV~\cite{Gleisberg:2003bi,Dittmaier:2002}.

\begin{figure}[htbp]
  \begin{center}
 \begin{equation*}
\parbox{35mm}{
\begin{fmfgraph*}(35,30)

\fmfleftn{l}{2}
\fmfrightn{r}{6}
\fmf{fermion,tension=3,la= $e^+$,la.si=left,la.di=0.15cm}{v1,l1}
\fmf{fermion,la= $\bar\nu_e$,la.si=left,la.di=0.15cm}{r1,v1}
\fmf{fermion,tension=3,la= $e^-$,la.si=left,la.di=0.15cm}{l2,v2}
\fmf{fermion,la= $e^-$,la.si=left,la.di=0.15cm}{v2,r6}
\fmf{photon,tension=1.5,label=$W^+$,la.di=0.1cm,la.si=left}{v1,i1}
\fmf{photon,tension=1.5,label=$\gamma/Z$,la.di=0.1cm,la.si=right}{v2,i2}
\fmf{fermion}{r5,i2,i1}
\fmf{fermion,tension=2,label= $t$,la.si=right,la.di=0.1cm}{i1,w1}
\fmf{fermion,tension=2}{w1,r2}
\fmffreeze
\fmf{photon,label=$W^+$,la.si=left,la.di=0.01cm}{w1,w2}
\fmf{fermion}{r3,w2,r4}
\fmfv{label= $b$,la.di=0.1cm}{r2}
\fmfv{label= $\nu_\mu$,la.di=0.1cm}{r4}
\fmfv{label= $\mu^+$,la.di=0.1cm}{r3}
\fmfv{label= $\bar b$,la.di=0.1cm}{r5}
\end{fmfgraph*}}
\qquad
\parbox{35mm}{
\begin{fmfgraph*}(35,25)
\fmfleftn{l}{2}
\fmfrightn{r}{6}
\fmf{fermion,tension=1}{v1,l1}
\fmf{fermion,tension=1}{l2,v1}
\fmf{photon,tension=1.5,label=$\gamma/Z$,la.di=0.1cm,la.si=right}{v1,i1}
\fmf{fermion,label= $t$,la.di=0.1cm}{i1,w2}
\fmf{fermion}{w2,r1}
\fmf{fermion}{r6,w1}
\fmf{fermion,label= $\bar t$,la.di=0.1cm}{w1,i1}
\fmffreeze
\fmf{photon}{w1,le1}
\fmf{photon}{w2,le2}
\fmf{fermion}{r2,le2,r3}
\fmf{fermion}{r5,le1,r4}
\fmfv{label= $\bar\nu_e$,la.di=0.1cm}{r5}
\fmfv{label= $b$,la.di=0.1cm}{r1}
\fmfv{label= $\nu_\mu$,la.di=0.1cm}{r3}
\fmfv{label= $\mu^+$,la.di=0.1cm}{r2}
\fmfv{label= $\bar b$,la.di=0.1cm}{r6}
\fmfv{label= $e^-$,la.di=0.1cm}{r4}
\end{fmfgraph*}}
\qquad
\parbox{35mm}{
\begin{fmfgraph*}(35,30)
\fmf{fermion,tension=2,la= $e^+$,la.si=left,la.di=0.15cm}{v 66,v 64}
\fmf{fermion,la= $\bar\nu_e$,la.si=left,la.di=0.15cm}{v  2,v 66}
\fmf{boson,lab=$W^+$,la.di=0.01cm,la.si=left}{v 66,v 78}
\fmfv{label=$\mu^+$,la.di=0.1cm}{v  4}
\fmf{fermion}{v  4,v 12}
\fmfv{label=$\nu_\mu$,la.di=0.1cm}{v  8}
\fmf{fermion}{v 12,v  8}
\fmf{boson}{v 78,v 12}
\fmfv{label=$b$,la.di=0.1cm}{v 16}
\fmf{fermion}{v 48,v 16}
\fmfv{label=$\bar{b}$,la.di=0.1cm}{v 32}
\fmf{fermion}{v 32,v 48}
\fmf{plain}{v78,v 48}
\fmf{boson,lab=$Z$,lab.si=right}{v 0,v78}
\fmf{fermion,la= $e^-$,la.si=left,la.di=0.15cm}{v  0,v  1}
\fmf{fermion,tension=2,la= $e^-$,la.si=left,la.di=0.15cm}{v128,v  0}
\fmfblob{5mm}{v78}
\fmfv{label=$H/Z$,label.angle=90,label.angle=120,label.di=0.1cm}{v48}
\fmfv{label=$W^+$,label.angle=-120,label.di=0.1cm}{v12}
\fmfleft{v 64,v128}
\fmfright{v  2,v  4,v  8,v 16,v 32,v  1}
\end{fmfgraph*}}
\end{equation*}
    \caption{Representative diagrams contributing to single top production, top pair production and vector boson fusion}
    \label{fig:single-top-back}
  \end{center}
\end{figure}
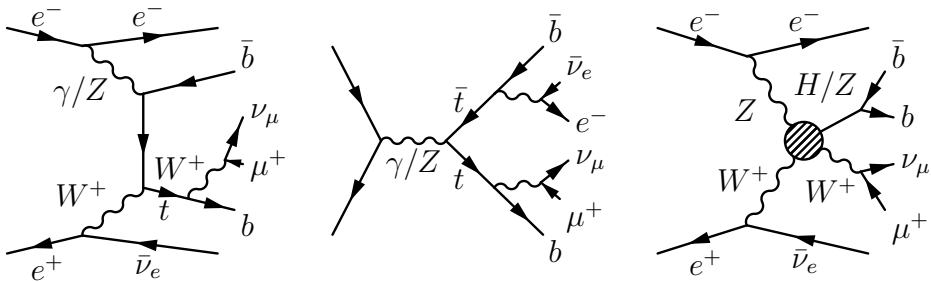

To obtain the single top signal from the cross section for the full
six fermion final state, 
contributions from  top pair production and 
vector boson fusion to the same final state have to be reduced (see
figure~\ref{fig:single-top-back}).
Top-pair
contributions are the dominant contributions already for the four
particle final state $e^+e^-\to t\bar b e^-\bar\nu_e$ and it 
 it has been suggested in~\cite{Boos:2001sj} either to impose
a cut on the invariant mass of the $e^-\bar\nu_e \bar b$
system
\begin{equation}\label{eq:top-mass-cut}
  |m_{\bar b,e^-\bar\nu_e}-m_t|>20\text{ GeV }
\end{equation}
or to eliminate the $s$-channel contributions altogether using
right-handed polarized electron and positron beams.  Both
prescriptions will adopted in our analysis.  Turning to six fermion
final states, there are additional contributions where no top quark is
produced at all, the main contribution being $Z$ and $H$ production by
vector boson fusion~(c.f. figure~\ref{fig:single-top-back}).  To
reduce this background, cuts on the invariant mass of the $b\bar b$
quark pair
\begin{equation}
  \label{eq:mass-cut}
|m_{\bar b b}-m_Z|>20\text{ GeV }\quad \text{and}\quad |m_{\bar b b}-m_H|>1 
\text{GeV}  
\end{equation}
will be used.
For definiteness, in the following the leptonic final state $e^+e^-\to
b\bar b\mu^-\bar\nu_\mu e^+\nu_e $ will be considered.  To appreciate
the size of the background without top production, in
table~\ref{tab:6ftop} the full cross section is compared to the result
for the process $e^+e^-\to t^* \bar b e^+\nu_e\to
b\mu^-\bar\nu_\mu\bar b e^+\nu_e $ with an intermediate off-shell top
quark.  Including only this top production subset of diagrams is not
compatible with gauge invariance since the $t$-channel top production
diagrams are connected to vector boson fusion diagrams by gauge
flips~\cite{Boos:1999}.  However, for the cut
$\theta_e>5^\circ$ on the scattering angle of the outgoing electron as
included in the cuts used in~\cite{Dittmaier:2002,Gleisberg:2003bi},
the numerical impact of this inconsistency appears to be small, at
least for the unpolarized case. The effect of the mass
cuts~\eqref{eq:mass-cut} on the full cross section and the top
production contribution is shown in table~\ref{tab:6ftop} for
unpolarized beams and for right-handed polarized electron and positron
beams where top pair production is suppressed.
Here $100\%$ polarization has been assumed for both beams, leaving the
discussion of more realistic polarization rates to future work.  Again, input
parameters and the remaining cuts are as in~\cite{Gleisberg:2003bi}.
\begin{table}[htbp]
  \begin{center}
    \begin{tabular}{|c|c||c|c||c|c| }
\hline
\multicolumn{6}{|c|}{ $\sigma(e^+e^-\to b\bar b e^+\nu_e\mu^-\bar\nu_\mu)$(fb)}\\\hline
$\sqrt s$& Cut & $ b\bar b e^+\nu_e\mu^-\bar\nu_\mu$  &
 $t^* \bar b e^+\nu_e$ &$ b\bar b e^+\nu_e\mu^-\bar\nu_\mu$ (RR) &
 $t^* \bar b e^+\nu_e$ (RR)   \\\hline
500&-&  5.831 (10)  &5.551 (9) &0.0455 (1)&0.0103 (1)   \\\hline
500&$m_{\bar b b}$ & 4.708 (9) &  4.600 (10)&0.0073 (4)&0.0074 (4)  \\\hline\hline
800&- & 3.150 (6) &  2.719 (4) &0.1714 (6)&0.0450 (1)    \\\hline
800& $m_{\bar b b}$  &2.639 (7)& 2.508 (7) &0.0335 (8)&0.0366 (9)  \\\hline\hline
2000&- & 1.461 (5) & 0.693 (2) &0.462 (2)&0.206 (1)    \\\hline
2000&$m_{\bar b b}$ & 0.691 (4) &  0.667 (4) &0.135 (5)&0.189 (6)   \\\hline
   \end{tabular}
\caption{Top production contribution and full cross section for  $e^+e^-\to b\bar b e^+\nu_e\mu^-\bar\nu_\mu$ before and after application of the mass
cut~\eqref{eq:mass-cut}.
          Fixed width, $\theta_e>5^\circ$, unitarity gauge }
\label{tab:6ftop}
  \end{center}
\end{table}

\begin{figure}[htbp]
 \begin{center}
      \includegraphics[width=.75\textwidth]{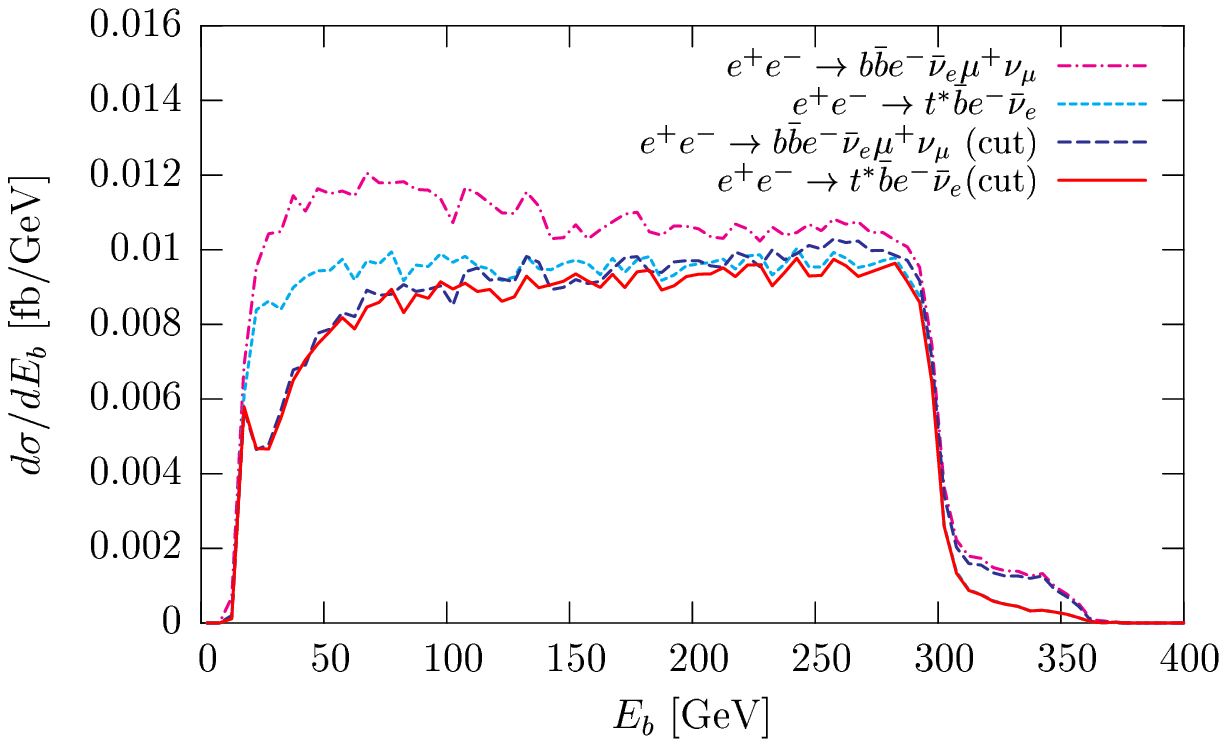}
      \includegraphics[width=.75\textwidth]{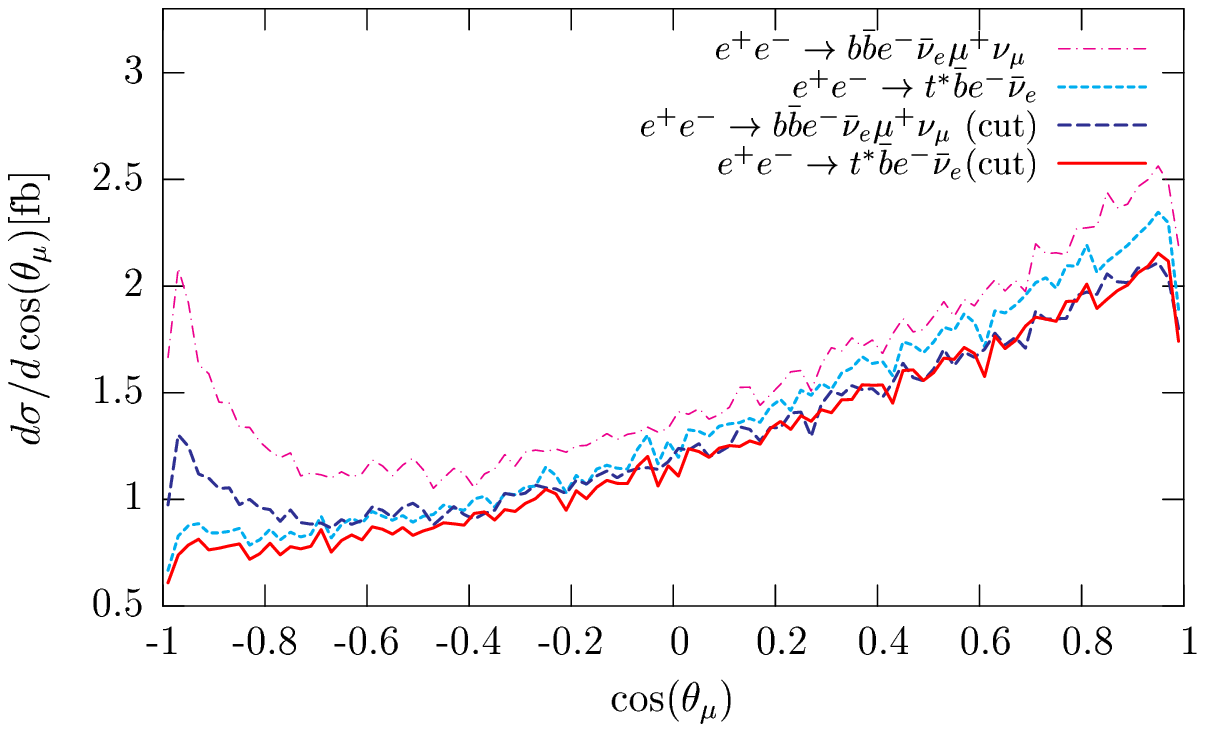}
\caption{Effects of a cut on the invariant mass of the bottom quark pair on
      the distribution of the energy of the $b$ quark (top) and the scattering angle of the the $\mu^+$ (below)  in $e^+e^-\to b\bar b e^+\nu_e\mu^-\bar\nu_\mu$ and $e^+e^-\to t^*
    \bar b e^+\nu_e$ at $800$ GeV.}
\label{fig:mass-cut}
\end{center}
\end{figure}

For unpolarized beams, the non-top production background is of the
order of $5\%$ for $\sqrt s=500$ GeV and becomes as large as the top
production contributions for $\sqrt s=2000$ GeV. After application of
the cuts~\eqref{eq:mass-cut}, a difference of $2-5\%$ remains.
 Effects of the mass cuts~\eqref{eq:mass-cut} on some
distributions of scattering angles and energies are shown in
figure~\ref{fig:mass-cut} for unpolarized beams and $\sqrt s =800$
GeV.  After applying the cuts in $m_{\bar b b}^2$, the distributions
for the full set of diagrams and the top-production subset in general
agree well.  For the scattering angle of the muon, there are some
deviations for backward scattering which might be relevant for the
measurement of anomalous couplings.  Such a distortion in the angular
distribution of the muon has also been observed by Yuasa {\it et.al.}
in~\cite{Accomando:1998} and in the
first reference of~\cite{Kolodziej:2001xe} in the
context of top-pair production.

For right-handed polarized beams the non-top contributions are of
larger importance than in the unpolarized case.  It should be observed
that starting at $\sqrt s =800$ GeV the result for the top-production
contribution exceeds the one from the full set of diagrams after the
cut on $m_{b\bar b}$, demonstrating the inconsistency of selecting the
top production subset.  For smaller scattering angles of the electron,
the selection of diagrams has found to be manifestly inconsistent
before imposing a cut on $m_{b\bar b}$.  For instance, for a cut on
the electron scattering angle of $\theta_e> 0.1^\circ$ and for $\sqrt
s=800$ GeV, the result for the top production subset exceeds the full
cross section by a factor of two for unpolarized beams and by an order
of magnitude for right-handed polarized beams.  Thus, from now always
the full set of Feynman diagrams will be included.

\begin{table}[htbp]
 \begin{center}
    \begin{tabular}{|c|c|c|c|c|c|}
\hline
\multicolumn{6}{|c|}{ $\sigma(e^+e^-\to b\bar b e^+\nu_e\mu^-\bar\nu_\mu)$(fb)}\\\hline
$\sqrt s$& gauge& Fixed Width &Complex Mass& Fudge Factor &Step Width \\\hline
500&UG  &5.913 (11)& 5.916 (15) & 5.832 (11) & 9.3 (1.9)  \\\hline
500&FG  &5.979 (25)& 5.925 (14) & 5.836 (12) & 11.57 (13)    \\\hline\hline
800&UG &3.541 (8)&3.549 (8)&3.528 (8)&4.46 (30) \\\hline
800&FG &4.984 (19) &3.543 (8) &3.527 (8) & 14.91 (14) \\\hline\hline
2000&UG &3.618 (16) &3.638 (14)&3.620 (20)&97.96 (44) \\\hline
2000&FG &4.955 (30) & 3.629 (17) &3.608 (18) &18.07 (9) \\\hline
   \end{tabular}
\caption{Comparison of  unitarity gauge (UG) and Feynman gauge (FG)  for different width schemes with  $\theta_e>0.01^\circ$}
\label{tab:6ftop-width}
 \end{center}
\end{table}
For smaller scattering angles of the electron---where the
contributions from single top production can be expected to be
large---care has to be taken also in the treatment of finite widths.
Compared to the related process of single $W$ production $e^+e^-\to u
\bar d e^+\nu_e$ where a consistent treatment of the $W$ width is
crucial for reliable results, in single top production also the finite
top width has to be treated carefully as described in
section~\ref{sec:omega}.  In table~\ref{tab:6ftop-width} the cross
sections in unitarity gauge and Feynman gauge are compared for
different schemes for the widths of gauge bosons and the top quark,
imposing a cut on the scattering angle of the electron of
$\theta(e^-)>0.01^\circ$.  Only in the complex mass and the fudge
factor scheme the results in unitarity gauge and Feynman gauge agree
within the errors from the Monte Carlo integration and numerically
stable results are obtained.  In unitarity gauge---where the fixed
width and the complex mass scheme differ in the top sector only by the
top-Higgs Yukawa coupling that is not relevant for the process under
consideration---also the results of these two schemes are consistent
with each other.  For a scattering angle $\theta_e> 5^\circ$ the
various schemes have been found to be consistent with each other, in
agreement with previous results~\cite{Dittmaier:2002}.

\begin{figure}[htbp]
  \begin{center}
 \includegraphics[width=.72\textwidth]{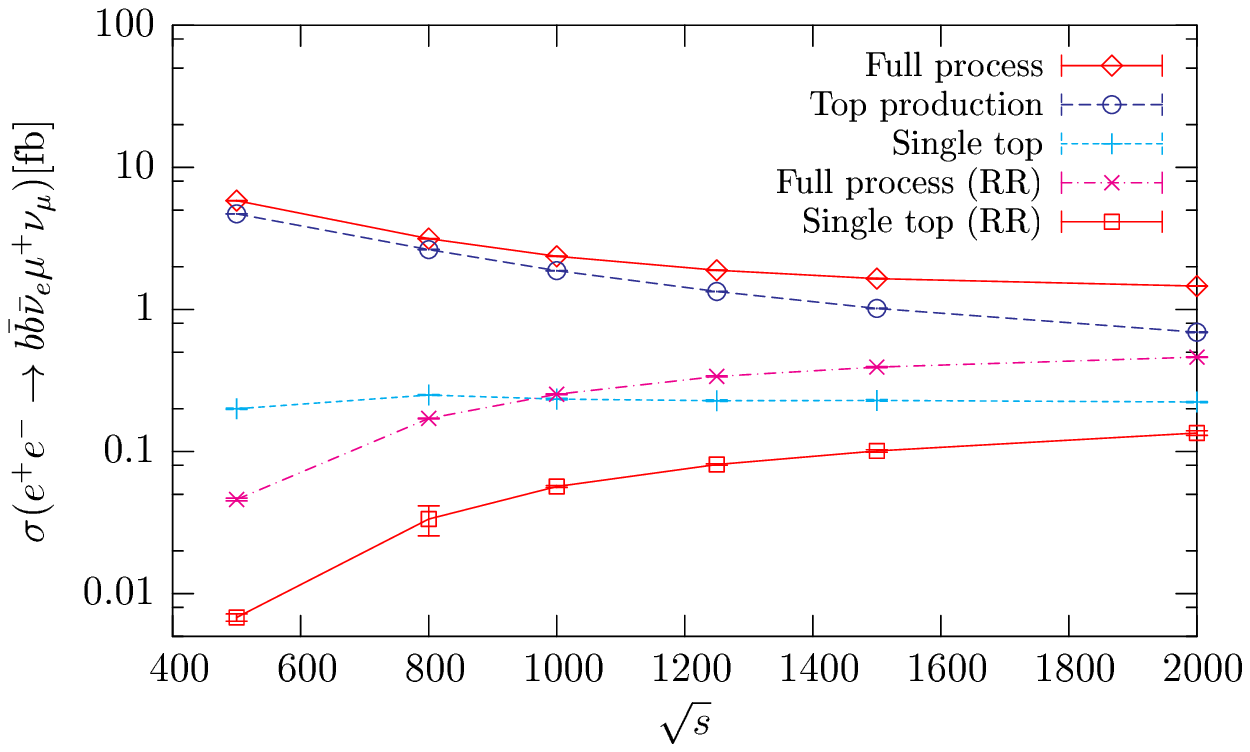}
 \includegraphics[width=.72\textwidth]{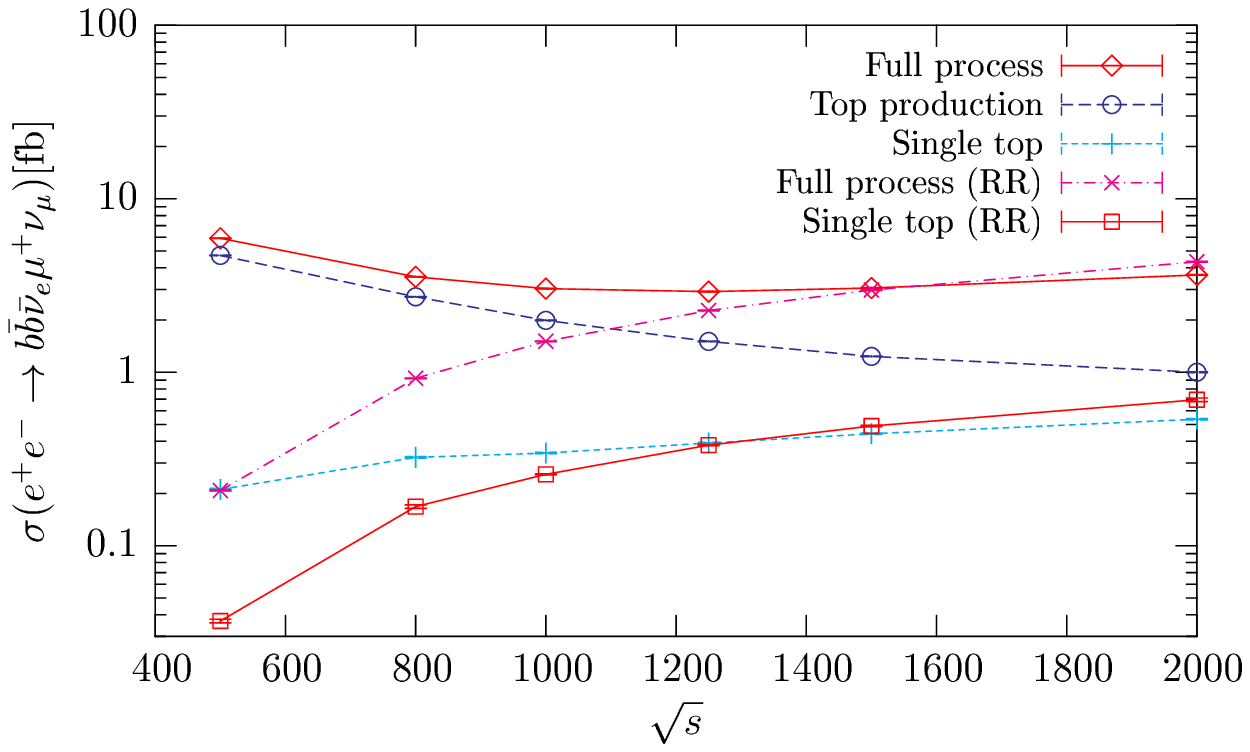}
 \caption{Contribution of single top and top production to the total cross section for $\theta_e >5^\circ$ (top) and   $\theta_e >0.01^\circ$ (below).
Here `top production'
denotes the cross section after applying the cut~\eqref{eq:mass-cut}
to reduce vector boson fusion while the single top cross section is
defined by the mass cut~\eqref{eq:top-mass-cut}. }
    \label{fig:sigma}
  \end{center}
\end{figure}
\begin{table}[htbp]
\begin{center}
\begin{tabular}{|c|c||c|c|}
\hline
\multicolumn{4}{|c|}{ $\sigma(e^+e^-\to b\bar b e^+\nu_e\mu^-\bar\nu_\mu)$(fb)}\\\hline
$\sqrt s$&$\theta_e$ &Unpolarized &RR \\\hline\hline
500&$5^\circ$ &0.200 (2) &0.0068 (4) \\\hline
500&$0.01^\circ$ &0.211 (4)  &0.0368 (9)   \\\hline\hline
800&$5^\circ$ &   0.250 (2) &0.0335 (8) \\\hline
800&$0.01^\circ$ & 0.323 (4)  & 0.168 (4) \\\hline\hline
2000&$5^\circ$ & 0.223 (2) & 0.135 (5)  \\\hline
2000&$0.01^\circ$ & 0.536 (4) &  0.694 (17) \\\hline
   \end{tabular}
\end{center}
\caption{Single top contribution to the  $b\bar b e^+\nu_e\mu^-\bar\nu_\mu$ final state
after performing the cuts~\eqref{eq:top-mass-cut} and~\eqref{eq:mass-cut}, Complex mass scheme}
  \label{tab:singletop_sigma}
\end{table}

We now turn to the results for the single top production contribution
to the process $e^+ e^-\to b\bar b e^-\bar \nu_e\mu^+\nu_\mu$.  The
results in the complex mass scheme are shown in
table~\ref{tab:singletop_sigma} for two different cuts on the electron
scattering angle.  In figure~\ref{fig:sigma} the cross sections are
plotted as a function of the energy.  Allowing smaller scattering
angles greatly enhances the polarized cross section, while the effect
on the unpolarized cross section is more moderate.  This is consistent
with the observation of~\cite{Boos:2001sj,Boos:1996vc} that the
$s$-channel subset of diagrams contributes considerably to single top
production for unpolarized beams so that forward scattering doesn't
dominate the cross section, in contrast to the related process of
single $W$ production.  For right-handed polarized beams, $s$-channel
diagrams don't contribute and forward scattering is the dominant
contribution.  As can be seen in figure~\ref{fig:sigma}, also the
vector boson fusion background forming the main difference between the
full process and the top production contribution grows for forward
scattering of the electron and with increasing energy.  It remains to
study the effects of anomalous couplings on cross sections and angular
and energy distributions. For semileptonic final states $b\bar b
e^-\bar \nu_e q\bar q'$ also QCD effects should be included.

\section{Associated Top-Higgs Production}
\label{sec:yukawa}
For Higgs boson with masses below $2 m_t$,  associated top Higgs production 
$e^+ e^-\to t\bar t H$ ~\cite{Djouadi:1992gp} is the most
promising process for measuring the top-Yukawa coupling.
Several studies on the relevant backgrounds and the achievable precision
of the measurement have appeared.
In~\cite{Moretti:1999kx}
the process $e^+ e^-\to H \bar b b W^+ W ^-$ and the subsequent decay
to semileptonic final states has been considered for a center of mass
energy of $500$ GeV, together with the background processes $e^+
e^-\to Z \bar b b W^+ W ^-$ and $e^+ e^-\to g \bar b b W^+ W ^-$.
In~\cite{Baer:1999} the process $e^+ e^-\to
t\bar t\, b\bar b$ and subsequent decays has been studied at
$500$ GeV and $1000$ GeV.  Experimental studies on the precision of
the measurement of the top quark Yukawa coupling have been presented
in~\cite{Juste:1999} and
anomalous couplings have been discussed in~\cite{Gunion:1996vv}.
The combined effects of
QCD~\cite{Dittmaier:1998} and
electroweak~\cite{You:2003zq}
radiative corrections 
have recently been computed and 
are of the order of $\sim 10\%$ for 800 GeV.

In the present section, I give results obtained with \texttt{O'Mega/WHI\-ZARD} for the
complete electroweak tree-level contributions to cross sections for
semileptonic and leptonic eight particle final states for energies up
to $2000$ GeV. For the final state $e^+ e^-\to b\bar b b\bar b W^+ W^-$ 
results on the QCD background obtained with \texttt{MadGraph}
and \texttt{WHI\-ZARD} are also included.
Again the same input parameters as in~\cite{Gleisberg:2003bi}
have been used, including $m_H=130$ GeV and $\Gamma_H = 0.429\times
10^{-2}\text{ GeV}$. As the only exception, for the bottom quark
mass---that also enters the Yukawa coupling---the value $m_b=3.0$ GeV
has been used to obtain a tree level branching ratio
$\mathrm{BR}(H\to b\bar b)\sim 0.53$ appropriate for $m_H=130$ GeV, in agreement with the result from
\texttt{HDECAY}~\cite{Djouadi:1998yw} where a running bottom
mass at the scale $m_H$ is used.  
For the strong coupling constant the value at the $Z$ pole $\alpha_s(m_Z)=0.118$ has been used. 
Results for different values of top and
Higgs masses will be considered in future work. The fixed width
scheme and the unitarity gauge have been used in the calculation.

For $m_H\lesssim 200$ GeV, the decay width of the Higgs boson is
smaller than that of the top quark or the $W$ boson so to see the
effects of including the full final state, we will first use the
approximation  of an on-shell  Higgs\footnote{This has been suggested to me by S.~Dittmaier}
and consider leptonic seven particle final states $H b\bar b \ell\bar
\nu_\ell \bar \ell' \nu_{\ell'}$. In table~\ref{tab:yukawatop7} the
full cross sections are compared to the contributions with an
intermediate top pair, i.e. $e^+ e^- \to H t^*\bar t^*\to H b\bar b \ell\bar
\nu_\ell \bar \ell' \nu_{\ell'}$  or an intermediate $W$ pair,
i.e. $e^+ e^- \to H b\bar b W^{+*} W^{-*}\to H b\bar b \ell\bar
\nu_\ell \bar \ell' \nu_{\ell'}$.
In the calculation, the same cuts as in~\cite{Dittmaier:2002,Gleisberg:2003bi} have been used; no cut has been imposed on the outgoing Higgs boson.
Going from the
subset of diagrams with an intermediate $t\bar t $ pair to that with
an intermediate $W$ pair amounts to an effect of $\lesssim 2\%$ for $800$ GeV
and $\sim 5\%$ for 2 TeV. The effect from including the complete fermionic final state
is small for energies up to $800$ GeV, but becomes important for higher energies and identical fermions in the final state.
 \begin{table}[!htbp]
  \begin{center}
\begin{tabular}{|c||c|c|c||c||c|}
\hline
\multicolumn{6}{|c|}{$\sigma (e^+ e^-\to H \bar b b \ell \bar \nu_\ell \bar\ell'\nu_{\ell'})$ (ab)} \\\hline
$\sqrt s$ &$\mu^-\bar\nu_\mu \tau ^+ \nu_\tau$ $(\bar t^* t^*)$  &$\mu^-\bar\nu_\mu \tau ^+ \nu_\tau $ ($W^{+*} W^{-*}$) &$\mu^-\bar\nu_\mu \tau^+ \nu_\tau $&$\mu^-\bar\nu_\mu e ^+ \nu_e $&$e^-\bar\nu_e e ^+ \nu_e $   \\\hline
500 & 1.450 (3) & 1.473 (3)  &1.467 (4)  & 1.465 (4)&1.468 (4) \\\hline
800 & 22.12 (3)& 22.57 (7)&22.57 (4) &22.51 (6) & 22.54 (5)  \\\hline
2000 &8.18 (1)  & 8.83 (6)&  8.76 (2)&9.35 (3) & 10.47 (13) \\\hline
   \end{tabular}
\caption{Cross sections for leptonic final states for associated top Higgs production with an on-shell Higgs boson ($m_H=130$ GeV) . Contributions with an intermediate top pair  and an intermediate $W$ pair are also shown.}
\label{tab:yukawatop7}
\end{center}
\end{table}

For $m_H\lesssim 140$ GeV, the Higgs decays predominantly to a pair of
$b$ quarks, so the experimental signature for associated top-Higgs
production in this mass region consists of four bottom quarks and the
decay products of the $W$ bosons. There are also background
contributions from associated top-$Z$ and top-gluon production with
the same final state~\cite{Moretti:1999kx}.  In the usual linear
parameterization of the scalar sector, gauge invariance requires to
include both Higgs and $Z$ boson exchange diagrams~\cite{Boos:1999}.
Since no forward scattering of electrons is involved, the numerical
effect of the inconsistency caused by including only the signal
diagrams is expected to be less significant than for single top
production, but this remains to be studied systematically.  To
appreciate the relation of the Higgs signal to the background from
associated $Z$ production, in table~\ref{tab:yukawatop6} the
electroweak contribution to the cross section for the six particle
final state $e^+ e^-\to b\bar b b\bar b\, W^+ W^-$ is shown, together
with contributions with an intermediate Higgs or $Z$ boson,
subsequently decaying to a $b\bar b$ pair.  Here no cuts have been
applied.  Adding up the contributions with an intermediate Higgs boson and
with an intermediate $Z$ boson, the difference to the full electroweak
cross section ranges from about $7\%$ at $\sqrt 500$ GeV to $2\%$ at
higher energies.
\begin{table}[htbp]
\begin{center}
\begin{tabular}{|c||c|c|c|}
\hline
\multicolumn{4}{|c|}{$\sigma_{\text{EW}} (e^+ e^-\to \bar b b b\bar b W^+ W^-)$ (fb)}\\\hline
$\sqrt s$& $H^* b \bar b W^{+} W^{-}$ &$Z^* b \bar b W^{+} W^{-}$& 
$b\bar b b \bar b W^{+} W^{-}$\\\hline
500 &0.0724 (3)  &  0.1423 (6)  & 0.2308 (5)  \\\hline
800&1.117 (3)  & 0.660 (2)  & 1.813 (3) \\\hline
2000 & 0.422 (8)  &0.501 (3)   & 0.940 (6) \\\hline
   \end{tabular}
\caption{Comparison of associated top Higgs production to $Z$ production backgrounds and the full set of diagrams contributing to the production of 4 bottom quarks and two $W$ bosons ($m_H=130 \mathrm{ GeV}$).}
\label{tab:yukawatop6}
\end{center}
\end{table}
To compare with~\cite{Moretti:1999kx}, these
results can be converted to cross sections for the semileptonic final
states by multiplying with the appropriate branching ratios.
For $\sqrt s=500$ GeV one obtains $\sigma(e^+e^-\to H b\bar b W^+
W^-\to b\bar b b\bar b \ell \bar \nu_\ell q_i\bar q_j)=0.0315 (1) fb$,
in good agreement with the result $0.033$ fb obtained
in~\cite{Moretti:1999kx} for 
$m_t=175$ GeV.  In agreement with~\cite{Baer:1999},
table~\ref{tab:yukawatop6} shows that for $m_H=130$ GeV at $\sqrt s=
800$ GeV the Higgs contribution exceeds the background from an
intermediate $Z$ boson.  

\begin{figure}[tbp]
  \begin{center}
 \includegraphics[width=.6\textwidth]{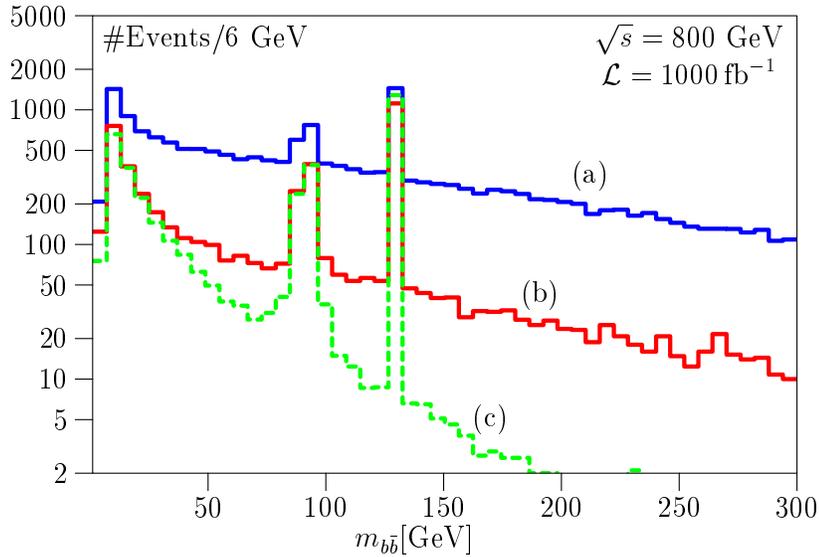}
\vspace{0.7cm}
 \caption{Invariant mass distribution of a $b\bar b$ quark pair for 
   combined electroweak and QCD contributions to \emph{(a)} $e^+ e^-\to
   b\bar b b\bar b W^+ W^-$ without cuts (blue), \emph{(b)} $e^+
   e^-\to b\bar b b\bar b W^+ W^-$ with a cut $|m_{Wb}-m_t|< 20$ GeV
   for the remaining $b$ quarks (red) and \emph{(c)} for $e^+
   e^-\to b\bar b t\bar t$ (green, dashed).}
    \label{fig:bb_tt}
  \end{center}
  \end{figure}

Apart from the electroweak background
contributions there is also a sizable QCD background
from gluon splitting $g\to b\bar b$.
  In in table~\ref{tab:yukawa_qcd}, we show the result obtained
  with the \texttt{O'Mega} matrix element for $e^+ e^-\to t\bar t
  g^*\to t\bar t b\bar b$, multiplied by an appropriate overall color factor
  that agrees well with the results obtained using
  the \texttt{MadGraph} matrix element in \texttt{WHI\-ZARD}.
\begin{table}[htbp]
\begin{center}
\begin{tabular}{|c||c|c||c|}
\hline
&\multicolumn{2}{c||}{$\sigma_{\text{QCD}}(e^+ e^-\to \bar b b t\bar t)$ (fb)} &$\sigma_{\text{QCD}}(e^+ e^-\to \bar b b b\bar b W^+ W^-)$ (fb) \\\hline
$\sqrt s$& \texttt{O'Mega}& \texttt{MadGraph} &\texttt{MadGraph}\\\hline
500 &1.072 (3)& 1.074 (3) &   2.602 (14)     \\\hline
800&2.197 (5)  & 2.219 (5)   & 3.106 (7)\\\hline
2000 & 1.306 (3) & 1.325 (4)  & 1.749 (7) \\\hline
   \end{tabular}
\caption{QCD background to associated top-Higgs production}
\label{tab:yukawa_qcd}
\end{center}
\end{table}
At $\sqrt s = 500$ GeV, an increase of the QCD cross section 
 by a factor of two has been observed in~\cite{Moretti:1999kx} in going
from the set of diagrams with the intermediate state  $t\bar t g^*$  to that with the
intermediate state $ b\bar b W^+W^- g^*$.
From table~\ref{tab:yukawa_qcd} we see that this effect
is even slightly larger for the full QCD contribution to the final state $b\bar b b \bar b W^+W^-$ that includes also diagrams where the gluon does not not split directly to a 
final state $b\bar b$ pair. For
higher energies the effect becomes smaller.
For a center of mass energy of $500$ GeV, cuts on the energy of the
$b$ quarks have been found efficient to reduce both the gluonic
background and that from $Z$ production~\cite{Moretti:1999kx} while
cuts on the invariant mass of the $b \bar b$ quark pairs have been
found more suitable for higher center of mass
energies~\cite{Baer:1999}, based on an analysis of the $b\bar b t\bar
t$ final state.  In figure~\ref{fig:bb_tt} we show the invariant mass
distribution of a given $b \bar b$ quark pair for the combined electroweak and QCD contributions to the $b\bar b t\bar t$ and the $b \bar b b\bar
b W^+ W^-$ final states.  In the second case, to reduce the background
we demand that the remaining $b$ quarks originate from top
decay by imposing a cut on the invariant mass of the $b W^+$ and $\bar
b W^-$ systems of $|m_{Wb}-m_t|< 20$ GeV. While 
these cuts are efficient in reducing both QCD and electroweak backgrounds,
the background is still considerably larger than for $e^+e^-\to b\bar b t\bar t$. These issues will be discussed further in future work.

 \begin{table}[bp]
  \begin{center}
\begin{tabular}{|c||c|c|c||c|c|c| }
\hline
&\multicolumn{3}{|c||}{$\sigma_{\text{EW}}(e^+ e^-\to b\bar b\bar b b \mu^- \bar \nu_\mu\tau^+\nu_\tau )$(ab)} 
&\multicolumn{3}{|c|}{$\sigma_{\text{EW}}(e^+ e^-\to b\bar b\bar b b \mu^- \bar \nu_\mu u\bar d)(ab) $} \\\hline
$\sqrt s$ &$ \sigma_{t^* t^*}$  & $\sigma_{W^{+*} W^{-*}} $ &$\sigma$  &$ \sigma_{t^* t^*}$  & $\sigma_{W^{+*} W^{-*}}$ &$\sigma$  \\\hline
500 &  2.289 (14)   & 2.337 (7)  &2.325 (8)  & 6.632 (38)   & 6.755 (25) & 6.779 (21)    \\\hline
800&18.69 (9)  & 18.96 (6) &18.97 (5)  &54.37 (25)&  55.30 (19)   & 55.14 (18)   \\\hline
2000 &8.68 (2) & 9.29 (12) & 9.28 (6) & 25.81 (7)  &27.95 (15)  & 27.54 (13) \\\hline
   \end{tabular}
\caption{Electroweak contributions to leptonic and semileptonic eight particle final states for
  associated top Higgs production 
together with the contributions with an intermediate top pair or an 
intermediate $W$ pair 
($m_H=130\mathrm{ GeV}$).}
\label{tab:8f}
\end{center}
\end{table}
Finally, table~\ref{tab:8f} shows the results for the full electroweak
contributions to the leptonic final state $e^+e^-\to b\bar b b\bar b
\tau^+ \bar\nu_\tau \mu^-\nu_\mu$ and the semileptonic final state
$e^+e^-\to b \bar b \bar b b \mu^-\bar\nu_\mu \bar u d$. Similar to
the process with an on-shell Higgs in table~\ref{tab:yukawatop7},
going from the subset of diagrams with an intermediate top pair to
that with an intermediate $W$ pair amounts to an effect of $\sim 2\%$
for $\sqrt s=800$ GeV, growing to $\sim 6\%$ at $\sqrt s =2$ TeV.
The inclusion of the full set of diagrams
contributing to the eight fermion final state shows no significant
effect for the final states without identical particles
considered here.

\section{Summary and Outlook} 
In this note, I have described results obtained with the 
computer programs \texttt{O'Mega} and \texttt{WHI\-ZARD} for six and eight fermion final 
states relevant for the measurement of the $V_{tb}$ CKM matrix element
and the Higgs-top Yukawa coupling at a linear collider.

In single top production, as discussed in
section~\ref{sec:single-top}, after applying cuts on the invariant
mass of the $b\bar b$ pair to reduce vector boson fusion backgrounds,
a difference between the top-production contribution and the full
cross section of the order of $5\%$ remains.  A small forward
scattering angle of the electron enhances the single top signal at
high energies, but also the vector boson fusion background.  For
this forward scattering of the electron, gauge invariant and numerically
stable results have only been obtained in the fudge factor and the
complex mass schemes for unstable particles.
%Since the latter
%prescription requires stable external particles, this gives further
%motivation to consider the full six fermion final states. 
A study of anomalous couplings has been left for
future work.

In the associated top-Higgs production process discussed in
section~\ref{sec:yukawa}, an effect of the order of $2-6\%$ has been
found in the electroweak cross section in going from the subset of diagrams with an intermediate top
pair to that with an intermediate $ b\bar b W^+ W^-$ state.
For the QCD background, the effect is much larger, as observed
already in~\cite{Moretti:1999kx}.  
The numerical effect of including the full set of
diagrams for the 8 fermion final states has turned out to be small if no
identical fermions are present in the final state.
Future studies will include a variation of the values of the top and
Higgs masses and the inclusion of anomalous Higgs-top couplings.
For heavier Higgs bosons  the decay modes
$H\to W^* W\to 4$ fermions should also be considered. 

\section*{Acknowledgements}
I thank Thorsten Ohl, Wolfgang Kilian and Andre van Hameren for
useful  discussions.
This work has been supported
by the Deutsche Forschungsgemeinschaft through the
Gra\-du\-ier\-ten\-kolleg `Eichtheorien' at Mainz University.

\providecommand{\href}[2]{#2}\begingroup\raggedright

\endgroup

\end{fmffile}
\end{document}